\documentclass[aps,prl,twocolumn,amsmath,amsart,amssymb,superscriptaddress]{revtex4}
\usepackage{graphicx}
\usepackage[T1]{fontenc}
\usepackage{amsmath}
\usepackage{color}
\usepackage{times}

\usepackage{ulem}

\newcommand{\BFCA}{Ba(Fe$_{1-x}$Co$_x)_2$As$_2$}
\newcommand{\BFA}{BaFe$_2$As$_2$}
\newcommand{\CFA}{CaFe$_2$As$_2$}

\begin{document}

%\modulolinenumbers[1]
%\linenumbers

%\title{Spin excitations of uniaxial-strain detwinned FeSe}
\title{Nematic quantum disordered state in FeSe}

\author{Ruixian Liu}
\affiliation{Center for Advanced Quantum Studies and Department of Physics, Beijing Normal University, Beijing 100875, China}

\author{Matthew B. Stone}
\affiliation{Neutron Scattering Division, Oak Ridge National Laboratory, Oak Ridge, Tennessee 37831, USA}

\author{Shang Gao}
\affiliation{Neutron Scattering Division, Oak Ridge National Laboratory, Oak Ridge, Tennessee 37831, USA}

\author{Mitsutaka Nakamura}
\affiliation{J-PARC Center, Japan Atomic Energy Agency (JAEA), Tokai, Ibaraki 319-1195, Japan}
\author{Kazuya Kamazawa}
\affiliation{Research Center for Neutron Science and Technology, Comprehensive Research Organization for Science and Society, Tokai, Ibaraki 319-1106, Japan}

\author{Aleksandra Krajewska}
\author{Helen C. Walker}
\affiliation{ISIS Facility, Rutherford Appleton Laboratory, STFC, Chilton, Didcot OX11 0QX, United Kingdom}

\author{Peng Cheng}
\author{Rong Yu}
\affiliation{Department of Physics, Renmin University of China, Beijing 100872, China}

\author{Qimiao Si}
\author{Pengcheng Dai}
\email{pdai@rice.edu}
\affiliation{Department of Physics and Astronomy, Rice Center for Quantum Materials, Rice University, Houston, TX 77005, USA}

\author{Xingye Lu}
\email{luxy@bnu.edu.cn}
\affiliation{Center for Advanced Quantum Studies and Department of Physics, Beijing Normal University, Beijing 100875, China}

\date{\today}

\begin{abstract}

The unusual quantum-disordered magnetic ground state intertwined with superconductivity and electronic nematicity in FeSe has been a research focus in iron-based superconductors. However, the intrinsic spin excitations across the entire Brillouin zone in detwinned FeSe, which forms the basis for a microscopic
understanding of the magnetic state and superconductivity, remain to be determined. Here, we use inelastic neutron scattering to map out the spin excitations of FeSe dewtinned with a uniaxial-strain device. We find that the stripe spin excitations (Q=(1, 0)/(0, 1)) exhibit the $C_2$ symmetry up to $E\approx120$ meV, while the N{\'e}el spin excitations (Q=(1, 1)) retain their $C_4$ symmetry in the nematic state. The temperature dependence of the difference in the spin excitations at Q=(1, 0) and (0, 1) for temperatures above the structural phase transition unambiguously shows the establishment of the nematic quantum disordered state. The similarity of the N\'eel excitations in FeSe and NaFeAs suggests that the N\'eel excitations are driven by the enhanced electron correlations in the $3d_{xy}$ orbital. By determining the key features of the stripe excitations and fitting their dispersions using a Heisenberg Hamiltonian with biquadratic interaction ($J_1$-$K$-$J_2$), we establish a spin-interaction phase diagram and conclude that FeSe is close to a crossover region between the antiferroquadrupolar, N\'eel, and stripe ordering regimes. The results provide an experimental basis for establishing a microscopic theoretical model to describe the origin and intertwining of the emergent orders in iron-based superconductors.

% The N\'eel spin excitations can be explained as they are close to the tri-critical point.

\end{abstract}

\maketitle
\noindent
The parent compounds of iron pnictide superconductors exhibit a tetragonal-to-orthorhombic structural phase transition at $T_s$ and then form collinear long-range 
antiferromagnetic (AF) stripe order at the wave vector $\mathbf{Q}_{AF}=(1,0)$ below $T_N$ ($T_N\leq T_s$) \cite{Scalapino_RMP,Dai_RMP}.  At temperatures below $T_s$, iron pnictides establish an electronic 
nematic phase where the electronic and magnetic properties along the $\mathbf{Q}=(1,0)$ direction are different from those along the $(0,1)$ direction
\cite{Fernandes2014, Si_NRM2016, Si_Hussey2023, Fernandes2022, Boehmer2022, Chu2010,Yi2011,Chu2012,Kuo2016,Lu14,Lu18}.
 With increasing electron or hole doping, both $T_N$ and $T_s$ decrease and vanish near optimal superconductivity but AF and nematic fluctuations remain, thus suggesting their important role for superconductivity \cite{Dai_RMP, Fernandes2014, Fernandes2022,Boehmer2022}. 
While most iron-based superconductors follow this paradigm, iron chalcogenide FeSe is different \cite{Hsu2008}.
With a simple layered structure, the stoichiomtric FeSe undergoes a tetragonal-to-orthorhombic structural (nematic) 
transition at $T_s=90$ K, and then becomes superconducting at $T_c\approx8$ K without static AF order (Fig. 1(a)) \cite{McQueen2009, Wang2016_1, Tanatar2016, Coldea2018}.  Since nematic phase transitions for iron pnictides and FeSe occur below room temperature with small
orthorhombic lattice distortions, samples are twinned below $T_s$ where the $(1,0)$ and $(0,1)$ directions are mixed and cannot be separated 
in most spectroscopic measurements \cite{Yi2011, Lu14}.  While
inelastic
 neutron scattering (INS) experiments on twinned iron pnictides 
such BaFe$_2$As$_2$ have spin excitations stemming from the stripe ordering $\mathbf{Q}_{AF}=(1,0)/(0,1)$ positions consistent with expectations \cite{Leland_BFA}, similar measurements on twinned FeSe reveal spin excitations at the stripe ($\mathbf{Q}_{AF}=(1,0)/(0,1)$) and N{\'e}el ($\mathbf{Q}_N=(1, 1)$) ordering wave vectors \cite{Wang2016_2}.
The absence of static AF order in FeSe was interpreted as arising from the magnetic frustration due to competing stripe and N{\'e}el 
interactions \cite{Wang2015,YuSi15,Glasbrenner15}.

\begin{figure*}
\includegraphics[width=16cm]{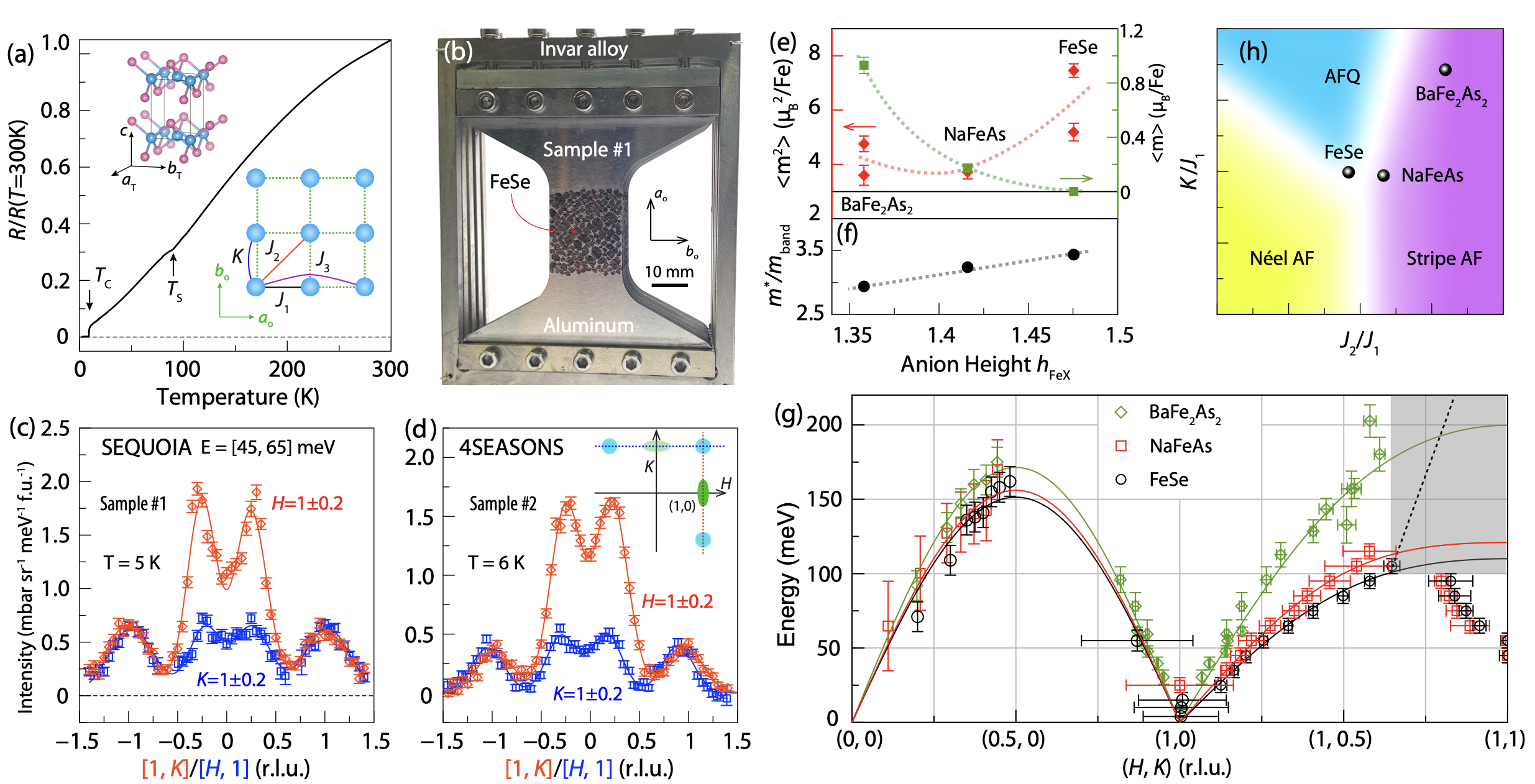}
\caption{{\bf Sample characterization, detwinning device and summary of the key results.} (a) Temperature-dependent resistivity of FeSe single crystal. The left-upper inset illustrates the crystal structure of FeSe in tetragonal notation; the right-lower inset shows the orthorhombic unit cell in the Fe plane. $J_1$, $J_2$, and $J_3$ represent the nearest-neighbor- (NN), next-NN, and next-next-NN exchange interactions between Fe ions. $K$ denotes the biquadratic interaction between the adjacent Fe ions. (b) The uniaxial-strain device based on the differential thermal expansion coefficients between the Invar alloy (Fe$_{0.64}$Ni$_{0.36}$) frame and the aluminum sheet.  (c), (d) One-dimensional constant-energy cuts ($E = 55\pm10$ meV) of the spin excitations along $[1, K]$ and $[H, 1]$ directions collected at the SEQUOIA (c) and 4SEASONS (d) time-of-flight spectrometers using incident energy of 147.5 meV and 80 meV, respectively. The inset of (d) depicts the positions of the stripe and N\'eel spin excitations in $[H, K]$ space, and the trajectories (blue and red dashed lines with arrowheads) for the 1D cuts shown in (c) and (d). 
(e), (f) Static (green squares) and fluctuating (red diamonds) magnetic moment (f), and the electron correlation ($m^*/m_{\rm band}$) at the $3d_{xy}$ orbital (f) as a function of the anion height $h_{\rm FeX}$. The data points in (f) are from ref. \cite{ZP11}. (g) Spin-excitation dispersions for {\BFA} (green diamonds), NaFeAs (red squares), and FeSe (black circles). The data points in the range (0, 0)-(0.5, 0) were measured with RIXS \cite{Lu22, Pelliciari2016}. The data points for {\BFA} and NaFeAs obtained with INS are extracted from refs. \cite{Lu18, Zhang2014}. The green, red, and black solid curves are the fittings of the dispersions with the $J_1$-$K$-$J_2$ model. The gray-shaded area marks the $(\mathbf{Q}, E)$ region where the spin excitations are heavily damped. The black dashed line schematically shows that the spin-excitation dispersion turns up and deviates from the fitting.
(h) The ratios $K/J_1$ and $J_2/J_1$ for the fittings of the dispersions for {\BFA} (ref. \cite{Lu18}), NaFeAs (ref. \cite{Zhang2014}), and FeSe.
The INS data of {\BFA}, NaFeAs, and FeSe were collected at $T = 7$ K, $5$ K, and $5$ K, respectively.}
\label{fig1}
\end{figure*}

As the broad spin excitations centered
around $\mathbf{Q}_N=(1, 1)$ merge with those stemming from $\mathbf{Q}_{AF}=(1,0)/(0,1)$ for $E\gtrsim70$ meV in a twinned sample \cite{Wang2016_2}, it is unclear how to untangle 
the intrinsic spin excitations of FeSe from those due to twin domains. To understand the unusual magnetic state of FeSe, it is therefore imperative to map out its spin excitations in a twin-free sample. In previous work, we have carried out INS experiments on partially detwinned FeSe by gluing them on uniaxial strained {\BFA} substrates 
\cite{Lu18,Chen19}. However, spin waves from the {\BFA} substrates overwhelm the magnetic signal from FeSe for spin excitation energies $E\gtrsim10$ meV and temperature 
across $T_s$ \cite{Chen19}. Resonant inelastic X-ray scattering (RIXS) experiments on the FeSe crystal detwinned by the same method are capable of avoiding the substrate 
spin waves. These measurements unveil a large spin excitation anisotropy
up to energy of $E\sim200$ meV below $T_s$ \cite{Lu22}; however, the limited momentum transfer of the Fe $L_3$ RIXS ($|\mathbf{q}_\shortparallel|\lesssim0.5 \frac{2\pi}{a_{\rm o}}$) means that one cannot access $\mathbf{Q}_{AF}$ and $\mathbf{Q}_N$ positions in these measurements.

In this work, we use INS to map out the energy-momentum dispersion of intrinsic spin excitations of FeSe, enabled by a 
newly designed low-background uniaxial-strain detwinning device (Fig. 1(b)) \cite{SI}, 
across the entire Brillouin zone (BZ). Our results identified the symmetries of the stripe and the N\'eel spin excitations (Figs. 1(c) and 1(d)), as well as their energy-momentum dispersions and the temperature dependence across the nematic phase transition. By analyzing the features of the spin excitations in FeSe, NaFeAs, and {\BFA}, we find that the N\'eel excitations are somewhat independent of the stripe excitations and should be driven by the enhanced electron correlations of the $d_{xy}$ orbital (Fig. 1(e)-(f)).
Through fitting the stripe-excitation dispersions using a Heisenberg Hamiltonian with 
biquadratic 
interaction ($J_1$-$K$-$J_2$) \cite{JK,RongYu2012},
 we establish a spin-interaction phase diagram for FeSe, NaFeAs and {\BFA}. As the low-energy stripe excitations in FeSe exhibit a linear-in-energy spectral weight and nematic spin correlations that can be described in an antiferroquadrupolar (AFQ) ordering 
 regime \cite{YuSi15,YuSi17},
 we argue that FeSe is positioned close to a crossover regime where the AFQ, N\'eel, and stripe orders intersect \cite{SI}.
The results provide an experimental basis for establishing a microscopic theoretical model to describe the origin and the intertwining of the emergent orders in iron-based superconductors (Figs. 1(f)-1(h)).

\begin{figure*}[htbp!]
\includegraphics[width=16 cm]{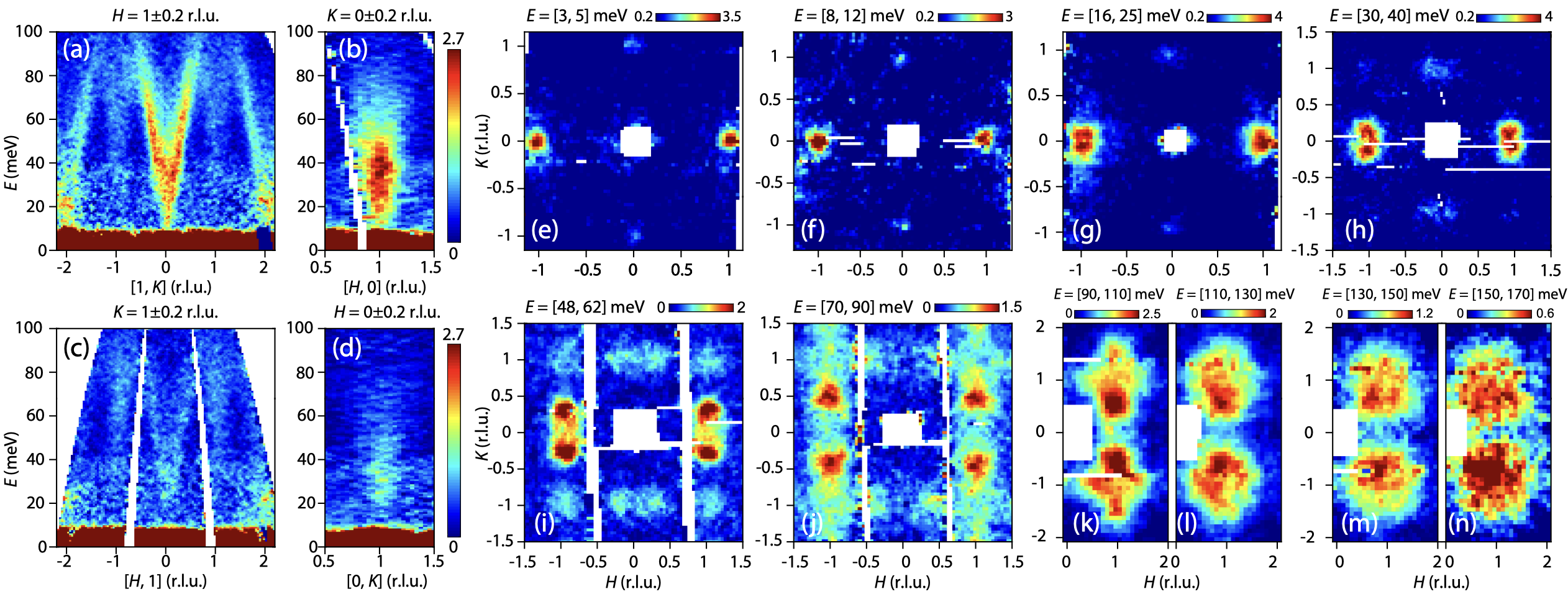}
\caption{{\bf Two-dimensional slices of the magnetic excitations in detwinned FeSe.} (a)-(d) Energy-vs-momentum slices of the magnetic excitations across $\mathbf{Q}=(1, 0)$, $(0, 1)$ and $(1, 1)$ measured with $E_i = 147.5$ meV at $T=5$ K. (e-n) Constant-energy slices of the magnetic excitations, with $E_i = 21$ meV (e), 36 meV (f)-(g), 80 meV (h), 147.5 meV (i)-(j), and 295 meV (k)-(n) at based temperatures. For the $E_i=147.5$ meV data ((a)-(d), and (i)-(j)), the scattering signal of an empty strain device with CYTOP has been subtracted as a background. A $|\mathbf{Q}|$-dependent background has been subtracted from the data shown in (e)-(h) \cite{Wang2016_2}, and energy-dependent flat backgrounds are subtracted from the data (k)-(n).
}
\label{fig2}
\end{figure*}

\hspace*{\fill}

\noindent
\textbf{Uniaxial-strain device to detwin a large amount of FeSe single crystals}

\noindent
We describe in Fig. 1(a)-(d) the characterization of the sample and the detwinning efficiency of the uniaxial-strain device. Figure 1(a) shows the resistivity of FeSe as a function of temperature. The two clear features on the resistivity curve (marked by black arrows) correspond to the nematic and the superconducting transitions at $T_s=90$ K and $T_c=8.5$ K, 
respectively. 
%The large residual resistivity ratio (RRR $\approx22$) indicates the high quality of our samples. 
Figure 1(b) displays a photo of Sample \#1 for the INS experiment, which consists of five separate uniaxial-strain devices stacked along the $c$-axis containing $\sim$1500 pieces ($\sim$1.61 grams) of thin FeSe crystals co-aligned along the tetragonal $[1,1,0]$ direction. The FeSe crystals are attached to $0.2$ mm thick aluminum alloy (6061) sheets using type-M CYTOP.
The uniaxial-strain device is designed based on the differential thermal expansion coefficients between the Invar alloy ($\alpha\approx-2\times10^{-6}$/K) outer frame and the aluminum ($\alpha\approx-24\times10^{-6}$/K) sheet fixed on the frame.
While cooling, the thermal expansion difference between the Invar-alloy frame and the aluminum sheet can generate a temperature-dependent anisotropic strain, reaching $\varepsilon=\varepsilon_V-\varepsilon_H\approx0.6\%$ at base temperature \cite{SI}, which is large enough to detwin FeSe with orthorhombic lattice distortion $\delta=[(a_{\rm o}-b_{\rm o})]/[(a_{\rm o}+b_{\rm o})]\approx0.27\%$ (corresponding to $\varepsilon\approx0.54$) at $T<<T_s$ \cite{Alan2016}. Neutron diffraction measurements on 26 pieces of FeSe crystals ($25$ mg) glued on one uniaxial-strain device reveals a detwinning ratio $\eta=\frac{P_1-P_2}{P_1+P_2}\approx71\%$ (for details see the Supplementary Information) \cite{SI}, where $P_1$ and $P_2$ represent the relative population of the two kinds of twin domains. For the samples used for INS experiments, we use the low-energy spin excitations at $E=4\pm1$ meV and  $10
\pm2$ meV to estimate the detwinning ratio and find the samples are highly detwinned with $\eta\approx58\%$ ($P_1/P_2\approx3.75$) (for details see the Supplementary Information) \cite{SI}.

\hspace*{\fill}

\noindent
\textbf{Spin excitations of detwinned FeSe at low temperature}

\noindent
%Having shown the detwinning strategy, we move to the INS experiments on the detwinned samples. 
Figure 1(c) shows one-dimensional (1D) constant-energy cuts with $E=55\pm10$ meV along the $[1, K]$ (red diamonds) and $[H, 1]$ (blue squares) directions measured on Sample \#1. The same cuts on Sample \#2 are presented in Fig. 1(d). 
In both panels,  the integrated intensity of the stripe excitations around $(1, 0)$ is $\sim3$ times of that around $(0, 1)$, indicating a large spin-excitation anisotropy. For comparison, the N\'eel excitations at $(\pm1,\pm1)$ are identical in lineshape and intensity. This confirms the $C_2$ symmetry of the stripe excitations reported in previous INS and RIXS studies \cite{Chen19, Lu22} and demonstrates that the N\'eel spin excitations are $C_4$ symmetric in the nematic state. 
%{\color{red}Add some discussions here.}

\begin{figure*}[htbp!]
\includegraphics[width=16cm]{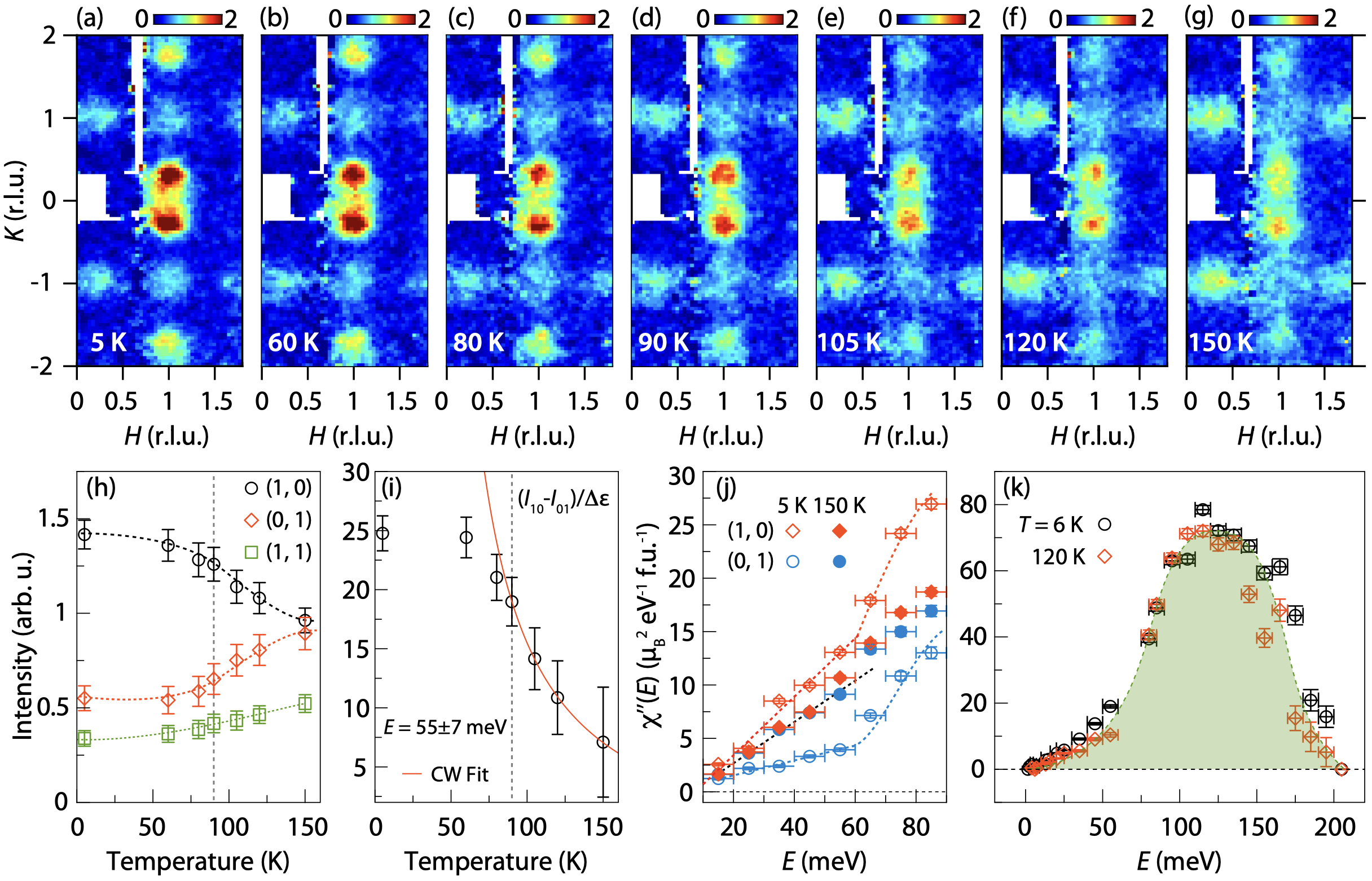}
\caption{{\bf Temperature dependence of the magnetic excitations in detwinned FeSe.} Constant energy slices of the magnetic excitations for detwinned FeSe single crystals with $E = 55\pm7$ meV at $T= 5$ K, $60$ K, $80$ K, $90$ K, $105$ K, $120$ K and $150$ K. The background scattering collected from an empty strain device with CYTOP has been subtracted from the data (a)-(g). (h) Temperature dependence of the integrated intensities of the magnetic excitations at {\bf Q} = (1, 0) (denoted by $I_{10}$), (0, 1) (denoted by $I_{01}$) and (1, 1) with $E = 55\pm7$ meV. Dashed curves are guides to the eyes. (i) Strain normalized nematic spin correlation $\Delta\psi(E, T)/\Delta\varepsilon=(I_{10}-I_{01})/\Delta\varepsilon$ with $E = 55\pm7$ meV. The red curve is a Curie-Weiss fitting of the data at $T = 90-150$ K with $\Delta\psi(E)/\Delta\varepsilon = \lambda/[a_0(T-T^*)+b_0]+\chi_0$. The vertical dashed lines in (h) and (i) mark the unstrained $T_s=90$ K.
(j) Comparison of local susceptibility $\chi''(E)$ (momentum averaged $\chi^{\prime\prime}({\bf Q}, E)$) for detwinned FeSe single crystals at {\bf Q} = (1, 0) and (0, 1), measured at $T = 5$ K (open symbols) and $150$ K (filled symbols). The data is corrected with the magnetic form factor and Bose factor. The dashed curves are guides to the eyes. (k) Energy dependence of local dynamic susceptibility $\chi''(E)$ for detwinned FeSe single crystals at $T = 6$ K (black open circles) and $120$ K (red open diamond). The horizontal and vertical error bars indicate the energy integration range for calculating $\chi''(E)$ and the statistical errors of one standard deviation. The data in (a)-(j) were collected on SEQUOIA with $E_i = 147.5$ meV. The data in (k) was collected on 4SEASONS with $E_i = 21, 36, 80$ and $295$ meV).}
\label{fig3}
\end{figure*}

Figure 2 reveals the energy and wave-vector dependence of the spin excitations of detwinned FeSe measured at base temperature. Figures 2(a)-2(d) show the spin excitations for $E_i=147.5$ meV projected onto $(\mathbf{Q}, E)$ planes with $\mathbf{Q}$ along the $[1, K], [H, 0], [H, 1],$ and $[0, K]$ directions, respectively. The scattering of an empty strain device (with CYTOP) measured under the same conditions has been subtracted from the data, leading to the clear spin excitation dispersions below $E=100$ meV in Figs. 2(a)-2(d) (see the Supplementary Information) \cite{SI}. Despite the absence of a stripe order, spin excitations arising from the stripe-order wavevector $(1, 0)$ are much stronger than the excitations emanating from the $(0, 1)$ and the $C_4$-symmetric N\'eel spin excitations at the $(\pm1, \pm1)$ positions (Figs. 2(a)-(d)). The energy-dispersion branches along the $[1, K]$ direction of the stripe excitations are very sharp (Fig. 2(a)),  while the stripe excitations in Figs. 2(b) and 2(d) are along the longitudinal $[H, 0]$ and $[0, K]$ direction appear non-dispersive and damped. 
Similar anisotropic damping of the stripe excitations between the $[1, K]$ and $[H, 0]$ directions was observed in {\CFA} \cite{Jun_CFA}, {\BFA} \cite{Lu18}, and NaFeAs \cite{Zhang2014}. While a clear dispersion along $[H, 0]$ can persist to $E\approx150$ meV in {\CFA} and $E\approx100$ meV for {\BFA}, it was damped so quickly that no dispersive feature can be resolved in NaFeAs. The highly anisotropic damping of the stripe excitations in FeSe is much akin to that in NaFeAs. As we will discuss later, the much stronger anisotropic damping in FeSe and NaFeAs could be attributed to the large anion height ($h_{\rm FeX}$) in these two materials that control the electron-correlation magnitude in the $d_{xy}$ 
orbital (Fig. 1(e)) \cite{ZP11,YuSi13}.

Figures 2(e)-2(n) are constant-energy intensity maps of the spin excitations in the $[H, K]$ plane, confirming again the $C_2$ and $C_4$ symmetry of the stripe and the N\'eel excitations, respectively. The stripe spin excitations at $(1, 0)$ are isotropic in momentum space at the low-energy range $E\lesssim15$ meV (Figs. 2(e)-(f)),
but exhibit anisotropic dispersion and damping at $E\gtrsim20$ meV. They propagate well along the $[1, K]$ direction but damp quickly along the $[H, 0]$ direction, consistent with the energy-momentum slices in Figs. 2(a)-2(d). The N\'eel excitations are visible at $E\gtrsim40$ meV (Figs. 2(h)-2(j)) and merge with the stripe excitations at $E\gtrsim90$ meV. The anisotropy between the stripe excitations at $(1, q_1)$ and $(q_2, 1)$ persists to $E=120\pm10$ meV and vanishes at higher energies, leaving four-fold symmetric broad scattering at around $(\pm1, \pm1)$ in Figs. 2(m) ($E=140\pm10$ meV) and 2(n) ($E=160\pm10$ meV).

The intensity difference between the spin excitations at $\mathbf{Q_1}=(1,q_1)$ and $\mathbf{Q_2}=(q_2,1)$ in the nematic state, termed nematic spin correlations, is defined as $\psi(E)=[S(\mathbf{Q_1}, E)-S(\mathbf{Q_2}, E)]/[S(\mathbf{Q_1}, E)+S(\mathbf{Q_2}, E)]$ (or defined via the imaginary part of the dynamic susceptibility $\chi^{\prime\prime}({\bf Q},E)$ in the same way) \cite{Lu18}. It was indirectly probed by RIXS in the limited region of the first BZ at the $\Gamma$ point \cite{Lu22}. Here we conclusively determine the energy scale of the nematic spin correlations as $E\approx120\pm10$ meV, roughly consistent with that ($E\approx200$ meV) determined in the RIXS study of detwinned FeSe \cite{Lu22}. 

\begin{figure*}[htbp!]
\includegraphics[width=16cm]{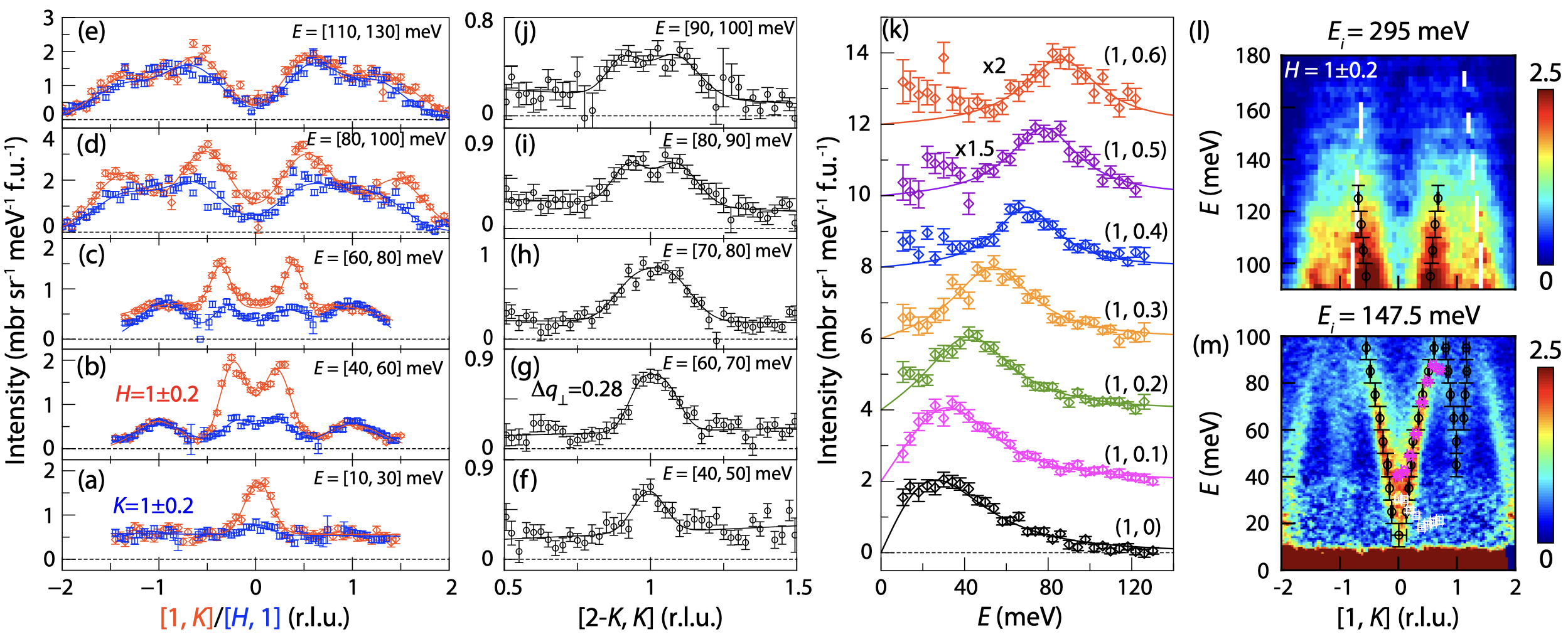}
\caption{{\bf One-dimensional cuts and energy dispersion of the magnetic excitations in detwinned FeSe.}  (a)-(j) Constant-energy cuts along the $[1, K]$, $[H, 1]$, and $[2-K, K]$ directions across the {\bf Q} = (1, 0) and (0, 1) ((a)-(e)) and {\bf Q} = (1, 1) ((f)-(j)) for detwinned FeSe single crystals at base temperature. The integral intervals perpendicular to the cut directions are $H=1\pm0.2$, $K=1\pm0.2$, and $\Delta q_{\perp}=0.28$ r.l.u., respectively. (k) Constant-Q cuts ($E_i = 147.5$ meV, $T = 5$ K) at $\mathbf{Q}=(1, q\pm0.05)$ ($q = 0, 0.1, ..., 0.6$). The data are fitted by a general damped harmonic oscillator model (solid curves). (l)-(m) Magnetic excitation dispersions along $[1, K]$ directions projected onto energy and momentum planes measured on (l) Sample \#2 with $E_i = 295$ meV, $T = 6$ K and (m) Sample \#1 with $E_i = 147.5$ meV, $T = 5$ K. The black circle symbols represent the magnetic excitation dispersions at {\bf Q} = (1, 0) and (1, 1), which are the Gaussian function fitting results of one-dimensional constant-energy cuts. }
\label{fig4}
\end{figure*}

\hspace*{\fill}

\noindent
\textbf{Temperature dependence of the spin excitations}

\noindent
The nematic spin correlation $\psi(E, T)$ is a manifestation of the electronic nematicity in the spin-spin correlation function and represents the nematic order parameter in the spin channel \cite{Fernandes2014}. Its temperature dependence across the nematic transition will provide more evidence concerning the origin of the nematic order.
Figures 3(a)-3(g) show the temperature dependence of the stripe and the N\'eel spin excitations in the energy range $E=55\pm7$ meV, measured at $T=5, 60, 80, 90, 105, 120$, 
and $150$ K. The integrated intensities as a function of temperature are shown in Fig. 3(h). 
The N\'eel excitations with smaller spectral weight increase slightly from $5$ K to $150$ K. 
On warming from $T=5$ K to $150$ K, the stripe excitations at $(1, 0)$ decrease, whereas the excitations at $(0, 1)$ increase gradually, leading to a diminishing $\psi(E)$.
No sudden change occurs for the stripe and the N\'eel excitations at the unstrained $T_s=90$ K, possibly because the structural transition is smeared out under the uniaxial strain \cite{Lu2016}. 
Note that the uniaxial strain decreases gradually with increasing temperature and could be reduced by $\sim40\%$ at $T=150$ K. 

The persistent $\psi(E)$ at $T>T_s$ induced by moderate uniaxial strain ($\varepsilon\sim0.3\%$) in the nematic fluctuating regime has been indirectly probed in a prior RIXS study of FeSe$_{1-x}$S$_x$ \cite{Liu2023}. Here, to obtain a quantitative understanding of $\psi(E, T\gtrsim T_s)$, we characterized the temperature-dependent uniaxial strain on a FeSe single crystal glued on a similar device using an optical method (see the Supplementary Information) \cite{SI} and got strain-normalized $\Delta\psi(E)/\Delta\varepsilon$=($I_{10}$-$I_{01})/\Delta\varepsilon$ for $E=55\pm7$ meV (Fig. 3(i)), which represents the nematic susceptibility in the spin (fluctuation) channel. The $\Delta\psi(E)/\Delta\varepsilon$ at $T\gtrsim T_s$ can be well described by a Curie-Weiss behavior $\Delta\psi(E)/\Delta\varepsilon = \lambda/[a_0(T-T^*)+b_0]+\chi_0$ (red curve in Fig. 3(i)), which generates a bare nematic transition temperature 
$T^*\approx34$ K.
This $T^*$ value is consistent with the Weiss temperature obtained by fitting the static nematic susceptibility derived 
from the elastoresistance measurements of FeSe \cite{Hosoi2016}. 
This further demonstrates that the electronic nematicity is driven by spin fluctuations \cite{Lu22, Liu2023}.

Figure 3(j) shows the temperature-dependent $\chi''(E)$ for the energy range $E\lesssim90$ meV measured on sample \#1 at SEQUOIA. The nematic stripe spin correlation ([$\chi''(\mathbf{Q_1}, E)-\chi''(\mathbf{Q_2}, E)]/[\chi''(\mathbf{Q_1}, E)+\chi''(\mathbf{Q_2}, E)$]) retains its magnitude at energies up to 90 meV at $T=5$ K and almost vanishes at $T=150$ K well above the unstrained $T_s$ under a moderate uniaxial strain (Fig. S6 in the Supplementary Information) \cite{SI}.

\hspace*{\fill}

\noindent
\textbf{Analysis of the spin-excitations in FeSe}

\noindent
A surprising discovery in the spin excitations is the linear-in-energy spectral weight (local dynamic susceptibility $\chi''(E)$) at the energy range $E\lesssim60$ meV measured at $T<<T_s$ and the temperature ($T=150$ K) well above $T_s$ (Figs. 3(j)-(k)). Some of us considered an $S=1$ generalized bilinear-biquadratic model on a square lattice and proposed that an $(\pi, 0)$ AFQ state could describe the magnetism in bulk FeSe \cite{YuSi15, YuSi17}. The linear energy dependence of the low-energy $\chi''(E)$ was an essential prediction/feature associated with the AFQ order, providing a clue to understanding the magnetic ground state, as we will discuss later.

Figure 3(k) shows the twin-domain averaged $\chi''(E)$ for the full energy range measured on Sample \#2 at 4SEASONS. 
%In comparison, the local dynamic susceptibility for the N\'eel excitations ($\chi_{11}''(E)$) in the energy range $E\approx[40, 90]$ meV is enhanced slightly at $T=150$ K, consistent with that reported in previous INS study \cite{Wang2016_2}. 
 It is consistent in lineshape with but shows ($\sim30\%$) higher peak intensity than that measured on a twined sample \cite{Wang2016_2}. The integral of the total spectral weight of $\chi''(E)$ generates the fluctuating moments $\big<m^2\big>=7.45\pm0.25 ~\mu_{\rm B}^2$ at $T=6$ K, and $6.23\pm0.34~ \mu_{\rm B}^2$ at $T=120$ K. Following the magnetic moment sum rule $\big<m^2\big>=g^2\mu_{\rm B}^2S(S+1)$ and with $g=2$ \cite{Dai_RMP}, we get $S\approx0.95\pm0.06$ for $T=6$ K and $S\approx0.84\pm0.08$ for $T=120$ K, corroborating the $S=1$ localized spin scenario for FeSe \cite{Wang2016_2}. %{\color{magenta}\sout{As the vanadium standard normalization (on different spectrometers) could result in $30\%$ uncertainty, the total fluctuating moments determined here is still consistent with the previous study on twinned sample }\cite{Wang2016_2}.}
 
\begin{figure*}[htbp!]
\includegraphics[width=12 cm]{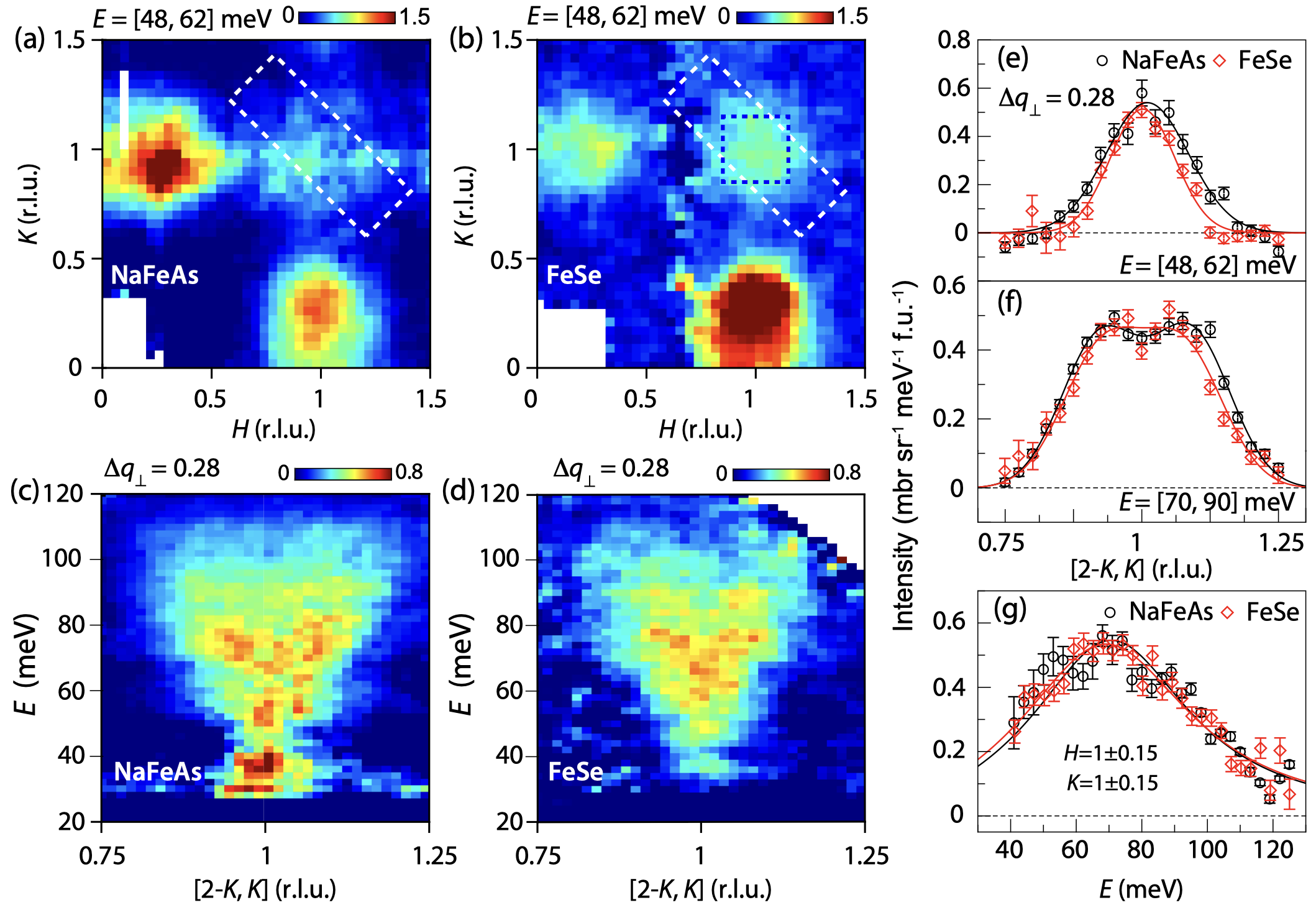}
\caption{\textbf{Comparison of the N\'eel excitations in NaFeAs and FeSe.} (a), (b) Constant-energy slices of the spin excitations in twinned NaFeAs ($E = 55\pm7$ meV) (a), collected with $E_i = 150$ meV and detwinned FeSe ($E = 55\pm7$ meV) (b), collected with $E_i = 147.5$ meV. (c), (d) spin-excitation dispersion along the transverse direction $[2-K, K]$ (dashed rectangles in (a)-(b)) across the $\mathbf{Q}$ = (1, 1) in twinned NaFeAs (c) and detwinned FeSe (d). (e), (f) Constant energy cuts along the $[2-K, K]$ direction for NaFeAs (black open circles) and FeSe (red open diamonds) with $E=55\pm7$ meV (e), and $E=80\pm10$ meV (f). The integral interval perpendicular to $[2-K, K]$ is $\Delta q_{\perp}=0.28$. (g) Energy cuts at $\mathbf{Q}=(1\pm0.15, 1\pm0.15)$ (marked by the blue dashed square in Fig 5(b)) for NaFeAs and FeSe.}
\label{fig5}
\end{figure*}

To obtain a quantitative understanding of the spin excitations, we plot in Figs. 4(a)-4(e) the constant-energy momentum cuts across (1, 0), (0, 1), and (1, 1) along the $[1, K]$ and $[H, 1]$ directions, with $E=20\pm10$, $50\pm10$, $70\pm10$, $90\pm10$, and $120\pm10$ meV, and in Figs. 4(f)-4(j) the momentum cuts along a diagonal $[2-K, K]$ direction with $E=45\pm5$, $65\pm5$, $75\pm5$, $85\pm5$, and $95\pm5$ meV.
The fit of the constant-energy momentum cuts with multi-Gaussian functions
generates the energy dispersions of the stripe and the N\'eel spin excitations, which are plotted onto the $E$-vs-$[1, K]$ slices in Figs. 4(l)-(m), where the stripe and the N\'eel excitations are resolved for energies up to $E\approx125$ meV.
We note that the N\'eel excitations touch the stripe excitation branch around $E\approx90$ meV while the latter persist to $E\approx130$ meV and dominate the spectral weight, indicating that they are indeed separate excitations arising from the competing stripe and N\'eel magnetic interactions \cite{Wang2016_2}.

In Fig. 4(k), we use a general damped harmonic oscillator function to describe
$\chi''(\mathbf{Q}, E)=A\,\frac{2\,\gamma\,EE_0}{\left(E^2-E_0^2\right)^2+(\gamma E\,)^2}$ and fit the energy cuts at $\mathbf{Q}=(1, q\pm0.05)$ ($q = 0, 0.1, ..., 0.6$) with $S(\mathbf{Q}, E)=\chi''(\mathbf{Q}, E)f^2(\mathbf{Q})$
(solid curves in Fig. 4(k)), where $E_0(q)$ is the undamped energy, $\gamma(q)/2$ is the damping rate, and $f(\mathbf{Q})$ the magnetic form factor of Fe$^{2+}$. The resulting damping rates $\gamma/2$ (white open squares in Fig. 4(m)) are much smaller than the undamped energies (magenta open diamonds in Fig. 4(l)), indicating the stripe excitations are only slightly damped for $q\le0.7$. This is 
unprecedented  
strong
evidence supporting the localized spin picture in FeSe. Thus, our results demonstrate that the $S=1$ localized spin picture is a correct starting point for describing the magnetism in FeSe, and it is likely that the magnetic state in FeSe is intimately connected with the AFQ regime.

\hspace*{\fill}

\noindent
\textbf{The N\'eel spin excitations}

\noindent
In iron-based superconductors, the N\'eel spin excitation at $(\pm1, \pm1)$ is not unique to FeSe, but had already been observed in NaFeAs hosting a static stripe order ($T_N\approx45<T_s=58$ K) with the ordered magnetic moment  $M=0.17 \pm 0.034 ~\mu_{\rm B}$/Fe and the total fluctuating moment $\big<m^2\big>=3.6 ~\mu_{\rm B}^2$ ($S\approx0.57$) \cite{Zhang2014,Tan2016,Scott2016}. 
Figures 5(a)-5(g) show the comparison of the N\'eel spin excitations in NaFeAs and FeSe, measured with $E_i=150$ meV on the ARCS spectrometer, and $E_i=147.5$ meV on the SEQUOIA spectrometer, respectively. 
%{\color{magenta}\sout{The incident energies are very close, allowing us to directly compare the background-subtracted magnetic scattering of FeSe with NaFeAs without considering the magnetic form factor.}} 
Figures 5(a) and 5(b) show magnetic excitations projected onto the $[H, K]$ plane with $E=55\pm7$ meV, in which similar N\'eel excitations at $(1, 1)$ are visible in both NaFeAs (Fig. 5(a)) and FeSe (Fig. 5(b)). The N\'eel excitations projected onto the $E$-vs-$[2-K, K]$ plane (Figs. 5(c) and 5(d)) also exhibit similar "V"-shaped dispersion. Furthermore, the constant-energy cuts across $(1, 1)$ along the $[2-K, K]$ direction at $E=55\pm7$ meV (Fig. 5(e)) and $E=80\pm10$ meV (Fig. 5(f)), and the energy cut at the $\mathbf{Q}$=(1, 1) show that the N\'eel excitations in NaFeAs and FeSe are almost identical. 

In addition, the stripe excitations of FeSe and NaFeAs exhibit similar anisotropic damping along the $[1, K]$ (transverse) and the $[H, 0]$ (longitudinal) directions, and similar much lower band top along the $[1, K]$ direction ($E\sim100$ meV) than {\BFA} ($E\sim200$ meV) \cite{Zhang2014, Scott2016}.
Figures 1(e) and 1(f) plot the systematic changes of the ordered moment ($\big<m\big>$), the fluctuating moment ($\big<m^2\big>$), and the electron correlation ($m^*/m_{\rm band}$) of these three compounds as a function of the anion height Fe-X (X=As, Se) ($h_{\rm FeX}$), suggesting
that the larger $h_{\rm FeX}$ in NaFeAs and FeSe (compared to {\BFA}) enhance the electron correlations in the $d_{xy}$ 
orbital \cite{Zhang2014,ZP11}. This has been used to explain small bandwidth along $[1, K]$ in FeSe and NaFeAs \cite{ZP11, Zhang2014, Scott2016}. As the N\'eel excitations are absent in {\BFA} and related materials, it is natural to speculate that the N\'eel excitations are also associated with the $d_{xy}$ orbital with enhanced electron correlations.

In electron-doped NaFe$_{1-x}$Co$_x$As,  while the low-energy stripe excitations associated with Fermi surface nesting evolve with the topology changes of the Fermi surfaces driven by electron doping, the relatively high energy ($E\gtrsim50$ meV) stripe excitations, as well as the N\'eel excitations, remain essentially unchanged across the whole superconducting regime ($x=0-0.11$) \cite{Scott2016}. These results suggest that superconductivity is coupled to the low-energy stripe excitations but not the N\'eel excitations. 
We note that the iron pnictogen height in ($h_{\rm FeAs}$) decreased by less than $1\%$ from $x=0$ to $0.11$ \cite{Parker2010}, consistent with the invariance of the N\'eel excitations.
In addition, several studies suggest that the stripe order could be ``restored'' in FeSe under $P\gtrsim1$ GPa hydrostatic pressure that reduces the anion height slightly \cite{Glasbrenner15, Alan2016, Yu_NMR, Cheng_HP}. Therefore,  the $h_{\rm FeX}$ could drive the ground state of FeSe across a phase boundary associated with the stripe order, and the pressured FeSe ($P\gtrsim1$ GPa) with stripe order fills the gap between FeSe and NaFeAs in Figs. 1(f) and 1(g).

\hspace*{\fill}

\noindent
\textbf{The spin-interaction phase diagram}

\noindent
To achieve a better understanding of the magnetic ground state in FeSe, we use a minimal $S=1$ $J_1$-$K$-$J_2$ model to fit the energy dispersions of the stripe spin excitations in FeSe, NaFeAs, and {\BFA} in Fig. 1(e). To improve the accuracy of the fittings, we include the dispersion near the $\Gamma$ point measured with RIXS  (data points between (0, 0) and (0, 0.5)) \cite{Lu22, Pelliciari2016}. The fitting of the dispersion for {\BFA} is a reference to show the validity of the fitting strategy. Previously, the fitting of the $S(\mathbf{Q}, E)$ in twinned {\BFA} with the $J_{1a}$-$J_{1b}$-$J_2$ (-$J_c$) model generates $J_{1a}=59.2$, $J_{1b}=-9.2$ and $J_2=13.6$ meV, corresponding to $J_1=25$ meV, $K=17.1$ meV ($J_{1a}$=$J_1$+2$K$, $J_{1b}$=$J_1$-2$K$), $J_2/J_1\approx0.54$ and $K/J_1\approx0.68$ in the $J$-$K$ model. In comparison, the fitting of the energy dispersion of {\BFA} in Fig. 1(e) gives $J_1=23.5$ meV, $K=16.1$ meV and $J_2=15$ meV ($J_2/J_1\approx0.64$ and $K/J_1\approx0.69$), agrees well with the fitting of the $S(\mathbf{Q}, E)$ in twinned {\BFA}. 

For FeSe (NaFeAs), the fitting of the energy dispersion provides $J_1=29.9 ~(28.8)$ meV, $K=11.9 ~(11.2)$ meV, and $J_2=11.0 ~(13.4)$ meV, leading to $J_2/J_1\approx0.37 ~(0.47)$ and $K/J_1\approx0.40 ~(0.39)$. The  $J_2/J_1$ of FeSe is close to that ($J_2/J_1$=0.413) reported in ref. \cite{Gu2022}. FeSe shows slightly larger $J_1$ and $K$ and smaller bandwidth than NaFeAs, further indicating the electron correlation in FeSe is slightly stronger than the other compounds (Fig. 1(f)). 

We plot in Fig. 1(h)
the trend in
 the magnetic interactions $J_2/J_1$ and $K/J_1$ of FeSe, NaFeAs, and {\BFA}. It is well known that {\BFA} is deep in the stripe-ordering region. NaFeAs hosting a weak stripe order accompanied by the N\'eel excitations should already be close to the cross-over between the stripe and the N\'eel regime. For FeSe, as the AFQ model can describe the linear-in-energy $\chi''(E)$ for $E\lesssim$ 60meV and $C_2$ symmetry of the stripe excitations, it should be near the AFQ ordering regime. Meanwhile, FeSe is close to NaFeAs in the phase diagram and exhibits also the N\'eel excitations. Thus, it should also be close to both the stripe and the N\'eel ordering regimes. All these key features can be qualitatively described in a zero-temperature phase diagram containing the stripe order, N\'eel order, and AFQ order generated by a $S=1$ bilinear-biquadratic $J_1$-$J_2$-$K$ model, as shown in Fig. 1(h) \cite{Hu2020}. We find that: (1) both $K/J_1$ and $J_2/J_1$ are essential in tuning the magnetic ground states; (2) $J_2/J_1$ plays a key role in driving the ground state from the stripe ordering region to the AFQ regime; and (3) FeSe is positioned close to a crossover regime where the AFQ, N\'eel, and stripe orders intersect \cite{Hu2020}.

\hspace*{\fill}

\noindent
In summary, our INS results on the uniaxial-strain detwinned FeSe clarify the symmetry of the stripe and the N\'eel spin excitations, characterize the nematic spin correlations, determine the magnetic interactions, and establish the evolution of the magnetic ground state in iron-based superconductors. The uniaxial-strain device suitable for INS developed in this work could also detwin or even apply uniaxial-strain on similar magnetic materials with layered structures.

\hspace*{\fill}

\noindent
{\bf Methods}

{\bf Sample preparation} The FeSe single crystals used in the present study were grown using the chemical vapor transport method. The direction of the self-cleaving edges of the FeSe crystals is tetragonal $[1,0,0]$,  as determined with a Laue camera. We co-aligned and glued $\sim1500$ pieces ($m_1\approx1.61$ grams) of thin FeSe crystals onto ten ($20\times22$ mm$^2$) areas (the front and back sides of 5 aluminum sheets) along the tetragonal $[1,1,0]$ direction using the hydrogen-free CYTOP, which were installed onto the invar alloy frames to form the Sample \#1 used in the neutron scattering experiments on SEQUOIA and MAPS time-of-flight spectrometers. We also prepared a Sample \#2 following the same way, which contained $\sim1000$ pieces ($m_2\approx1.24$ grams) of thin FeSe crystals and was used in the neutron scattering experiment on the Fermi chopper spectrometer 4SEASONS.

{\bf Neutron scattering experiments}  
The INS experiments were performed on the SEQUOIA \cite{Sequoia}, 4SEASONS \cite{4Seasons}, and MAPS \cite{MAPS} time-of-flight spectrometers at the Spallation Neutron Source at the Oak Ridge National Laboratory (ORNL), Materials and Life Science Experimental Facility (MLF) at the Japan Proton Accelerator Research Complex (J-PARC), and the ISIS spallation neutron source, Rutherford Appleton Laboratory (RAL), respectively. 
We defined the wave vector $\mathbf{Q}$ in reciprocal space as $\mathbf{Q} = H\mathbf{a^*} + K\mathbf{b^*} + L\mathbf{c^*}$, where $H, K, L$ are Miller indices and $a^*$ = {\bf \^a}~$2\pi/a_o$, $\mathbf{b^*}$ = {\bf\^b}~$2\pi/b_o$, and $\mathbf{c^*}$ = {\bf\^c}~$2\pi/c$ are reciprocal lattice unit (r.l.u.) vectors with $a_{\rm o}\approx5.33$ \AA, $b_{\rm o}\approx5.31$ \AA ~and $c\approx5.49$ \AA.

%The scattering intensity obtained from the INS experiments $\frac{d^2\sigma}{d{\mathbf{\Omega}}dE}\frac{k_i}{k_f}$ contains magnetic scattering signal and an background 

{\bf Author contributions}

X.L. conceived this project and designed the uniaxial-strain device. R.L. grew and coaligned the crystals. R.L., M.B.S., S.G. M.N., K.K., A.K., H.C.W., and X.L. performed the inelastic neutron scattering experiments. R.L., P.C., and X.L. performed the neutron diffraction measurements. R.L. and X.L. analysed the data. R.Y. and Q.S. carried out theoretical and computational analyses.
X.L. wrote the manuscript with inputs from R.L, P.D., R.Y., and Q.S. All authors made comments.

{\bf Acknowledgement}

The work at Beijing Normal University is supported by National Key Projects for Research and Development of China with Grant No. 2021YFA1400400 and the National Natural Science Foundation of China (grants nos. 12174029, and 11922402) (X.L.). 
The work at Renmin University of China is supported by the National Science Foundation of China Grant No. 12174441.
P.D. is supported by the U.S. DOE, BES under grant no. DESC0012311. 
Q.S. is primarily supported by the U.S. DOE, BES under Award No. DE-SC0018197, 
and by the Robert A. Welch Foundation Grant No. C-1411.
This research used resources at Spallation Neutron Source, a U.S. DOE Office of Science User Facility operated by ORNL. We acknowledge the neutron beam time from J-PARC with Proposal No. 2019A0002. We gratefully acknowledge the Science and Technology Facilities Council (STFC) for access to neutron beam time at ISIS \cite{ISIS_beamtime}.
%-------------------------------------------------------------------------------------

%------------------------------------------------------------------------------------

\end{document}